\documentclass[epj]{svjour}
\usepackage{graphicx}
\usepackage{dcolumn}
\usepackage{bm}
\usepackage{wrapfig}
\usepackage[T1]{fontenc}
\usepackage{textcomp}
\begin{document}
\title{The Phase Behavior of Mixed Lipid Membranes in Presence of 
the Rippled Phase}
\titlerunning{The Phase Behavior of Mixed Lipid Membranes}
\author{N. Shimokawa\inst{1}, S. Komura\inst{1} \and
D. Andelman\inst{2}
}                     
%
%
\institute{
Department of Chemistry,
Graduate School of Science and Engineering,
Tokyo Metropolitan University, Tokyo 192-0397, Japan
\and
School of Physics and Astronomy, Raymond and Beverly Sackler Faculty
of Exact Sciences, Tel Aviv University, Ramat Aviv 69978, Tel Aviv,
Israel
}
\date{Received: date / Revised version: date}
%
\abstract{
We propose a model describing liquid-solid phase coexistence in mixed
lipid membranes by including explicitly the occurrence of a
rippled phase.
For a single component membrane, we employ a previous model in
which the membrane thickness is used as an order parameter.
As function of temperature, this model properly accounts for the 
phase behavior of the three possible membrane phases: solid, liquid 
and the rippled phase. 
Our primary aim is to explore extensions of this model to binary 
lipid mixtures by considering the composition dependence of important 
model parameters.
The obtained phase diagrams  show various liquid, solid and rippled 
phase coexistence regions, and are in quantitative agreement with 
the experimental ones for some specific lipid mixtures.
} 
\maketitle

\section{Introduction}
\label{sec:intro}

In recent years, domain formation in biomembranes and artificial
membranes has attracted great attention in connection to the so-called
``raft formation'' in biological cell membranes \cite{SI97}.
It is believed that the appearance of such domains (rafts) in membranes
plays an important role for various cell functions
\cite{SI00,Engelman}.
Beside their interest in biocellular processes, raft formation has a
fundamental physical interest, because it offers an special example
of two-dimensional phase separations coupled to internal and external
membrane degrees of freedoms \cite{LA}.
Using fluorescence microscopy or X-ray diffraction techniques,
the lateral phase separation between different liquid phases, or
between the solid and liquid phases has been observed
for ternary-component vesicles consisting of saturated lipid,
unsaturated lipid, and cholesterol \cite{VK1,BHW,VK2,VK3,London}.
Although the biological significance of rafts is not yet fully 
understood, it serves as our primary motivation to explore, from a 
physical point of view, phase transitions, domains and phase 
coexistence in multi-component lipid membranes.

Lateral phase separation in membranes occurs even for the simpler case
of binary lipid mixtures without any added cholesterol.
The coexistence is between liquid-like ($L_{\alpha}$) and solid-like
($L_{\beta'}$ called ``gel'') phases, and it can be visualized using
techniques such as two-photon fluorescence microscopy \cite{BG}.
For example, for DPPC/DPPE lipid mixture, the solid domains
exhibit morphologies such as hexagonal, dumbbell or dendritic
shapes \cite{BG}, while for  DPPC/DLPC lipid vesicle the solid domains
appear to be stripe-like and are monodispersed in their domain width
\cite{BGZEP}. Shapes of solid domains found on spherical fluid vesicles 
has been also addressed theoretically \cite{SG}.
By assuming the additivity of  stretching and line
energies,  the model predicts a phase diagram including cap, ring 
and ribbon phase domains on a spherical surface.

Although visualization of domains embedded in fluid membranes has
become possible only recently, the phase diagram of various binary
lipid mixtures in bilayer membranes has been known for some time 
\cite{WM,ALG}. 
In a previous model investigated by some of us \cite{KSOA,KSO}, a
coupling between composition and internal membrane structure was
proposed and the resulting phase diagram was calculated.
It includes a coexistence region between $L_{\alpha}$ and $L_{\beta'}$
phases in agreement with experiments on specific lipid mixtures.
However, for some  lipids, such as DMPC or DPPC, another
distinct solid-like phase is known to occur  and is called the
``rippled''  or $P_{\beta'}$ phase.
The unique feature of this phase is that the membrane shape is
spatially modulated while the lipid hydrocarbon tails are ordered.
The specific phase diagrams of DMPC/DPPS, DPPC/DPPS
\cite{LMcC}, or DPPC/DPPE \cite{BWGG} mixtures have been explored in 
experiments, and show the three phases
($L_{\alpha}$, $L_{\beta'}$, $P_{\beta'}$) together with
the coexistence regions between them. However, these phase diagrams
have not yet been considered theoretically.

The main objective of the present paper is to provide a simple
model that describes the liquid-solid coexistence in binary lipid
mixtures while considering the possibility of an intermediate rippled 
phase.
Our starting point is a model introduced by Goldstein and
Leibler (GL) that accounts for the succession of phase transitions
as applied to single-component lipid membrane \cite{GL1,GL2}.
In the GL model, the membrane thickness is used as a scalar order
parameter.
We combined some of the ideas presented in the GL model with our 
own previous model for binary (or even ternary) lipid 
mixtures~\cite{KSOA,KSO}.
For binary lipid mixtures, the model parameters are taken to be 
dependent on the relative lipid composition, and resulting in
several types of phase diagrams that are in quantitative agreement 
with experiments on specific lipid mixtures.

In the next section, we first review the GL model describing the
phase transitions in a single-component lipid membrane.
The corresponding mean-field phase diagram is obtained in terms of
the temperature and membrane elastic constants.
In Sec.~\ref{sec:binary_model}, we propose an extension of
the GL model for a two-component lipid membrane, and calculate the
binary phase diagrams.
Some discussions and comparison to other models are discussed in
Sec.~\ref{sec:discussion}.

\section{Single-component lipid system}
\label{sec:single}

In a single-component lipid bilayer, one typically observes a
discontinuous first-order phase transition from the $L_{\alpha}$ 
to the $L_{\beta'}$ phase as the temperature is decreased.
This is  called the ``\textit{main transition}'' and is associated 
with ordering of the hydrocarbon tails of the lipid molecules.
In the $L_{\alpha}$ phase, the hydrocarbon chains are disordered (and
the phase is liquid-like), while they order by stretching and tilting
in the $L_{\beta'}$ (solid-like) phase.
Several theoretical models have been proposed in the past to describe
the main transition in isolated membranes \cite{Doniach1,IA}.

In other single-component lipid systems, however, a rippled
($P_{\beta'}$) phase appears between the $L_{\alpha}$ and
$L_{\beta'}$ phases for stacked bilayer membranes \cite{ZSGEH,WW,HR}.
This phase is more peculiar and many studies have been devoted to
understand better its modulation wavelength and amplitude
\cite{CBWWB}.
For a stack of lipid bilayers in water (a lamellar phase), the phase
diagram as a function of relative humidity and temperature was 
reported \cite{SSSPC,SSSC}.
Several attempts have been made to describe the rippled phase
theoretically such as those based on a molecular level description
\cite{Doniach2,CS}, or Monte Carlo simulations \cite{MS}.
In continuum theories of the rippled phase, different quantities have
been suggested for the order parameter.
Examples are the membrane thickness \cite{GL1,MFLM} or the
configuration of the hydrocarbon chains (ratio of the trans bonds
in the chains) \cite{HK}. In both cases the order parameter is scalar.
More recently, a Ginzburg-Landau theory was proposed by Chen, Lubensky,
and MacKintosh (CLM) who employed a vectorial order parameter 
representing the tilt of the lipid molecules \cite{LM,CLM}.

We further discuss the CLM model in Sec.~\ref{sec:discussion} below,
but our own starting point is based on a model proposed by Goldstein
and Leibler (GL) for single-component lipid membranes \cite{GL1,GL2}.
To describe the main transition involving chain ordering and
stiffening, GL introduced a dimensionless  scalar order parameter 
of the membrane, $m(\mathbf{r})$:
\begin{equation}
m(\mathbf{r})=\frac{\delta(\mathbf{r})-\delta_0}{\delta_0}.
\end{equation}
This parameter depends on the actual membrane thickness
$\delta(\mathbf{r})$, and on the constant membrane thickness
$\delta_0$ of the $L_{\alpha}$ phase.
The two-dimensional lateral position within the bilayer plane is
denoted by $\mathbf{r}$.
Notice that $m$ encapsulates changes that may occur in several degrees
of freedom, including, for example, the conformations of the hydrocarbon
chains, molecular tilt and positional ordering.
In the present paper, the bilayer nature of the membrane is not taken
into account and the bilayer thickness is simply taken as a sum of the 
two monolayer thicknesses.
The role of the bilayer structure in the formation of rippled phases
was explicitly considered in previous models \cite{KK,SSN}.

For simplicity, we assume that the rippled phase is spatially
modulated only in the $x$-direction, and ignore two-dimensional
rippled phases such as the square lattice phase \cite{LM,CLM} or
the hexagonal phase \cite{CM1,CM2}.
For an isolated lipid bilayer membrane, the stretching free energy
per lipid molecule is \cite{GL1}
\begin{eqnarray}
f_{\rm st} & = & \frac{1}{2}a_2 m^2 + \frac{1}{3}
a_3 m^3 + \frac{1}{4} a_4 m^4
\nonumber \\
&+ &
\frac{1}{2} C \left( \frac{{\rm d}m}{{\rm d}x} \right)^2 +
\frac{1}{2} D \left( \frac{{\rm d}^2 m}{{\rm d} x^2} \right)^2.
\label{free_energy}
\end{eqnarray}
A very similar free energy was proposed by Marder \textit{et al.}
\cite{MFLM}.
The first three terms are the Landau expansion in powers of the order
parameter $m$.
Only $a_2$, the second order term coefficient, has an explicit 
temperature dependence: $a_2 = a_2'(T-T^{\ast})$, with $T^{\ast}$ 
being a reference temperature. 
(It actually is the critical temperature in the absence of the cubic 
term.)
Because the $L_{\beta'} \rightarrow L_{\alpha}$ phase transition 
is known to be first-order, $a_3$ is taken to be negative, whereas 
$a_4$ is always positive to ensure stability of this free energy 
expansion.
The next two terms are related to the lowest order gradients of $m$.
(Note that  the $x$-coordinate is rescaled in units of a molecular 
length scale $\ell$ so that $x \to x/\ell$ is dimensionless.) 
These gradient terms represent elastic out-of-plane undulations of 
the membrane and their coefficients are the elastic constants $C$ 
and $D$, respectively.
The coefficient $C$ can be either positive or negative,
but $D$ is always positive
in order to ensure stability of the expansion.
The physical origin of a negative $C$ value can be related to the 
coupling between the conformation of the chains and the curvature of the
lipid/water interface, or to interactions between the polar head-groups
of the lipids and water \cite{GL1}.
This point will be further discussed later in Sec.~\ref{sec:discussion}.
In the absence of spatial gradient terms in Eq.~(\ref{free_energy})
and if $m(x)$ is constant in space, the main transition temperature 
$T_{\rm m}$ is calculated from the conditions 
$f_{\rm st}={\rm d}f_{\rm st}/{\rm d}m=0$. 
It is related to $T^\ast$ by
\begin{equation}
T_{\rm m}= T^{\ast} + \frac{2a_3^2}{9a'_2a_4}.
\label{main}
\end{equation}
With nonzero $a_3$, a first-order melting transition at $T_{\rm m}$ 
preempts the critical point at $T^*$.

When $C>0$, the equilibrium phase is always homogeneous in space;
either as a $L_{\beta'}$ or $L_{\alpha}$ phase.
On the other hand, for $C<0$, the rippled phase with a characteristic
wave vector $q^{\ast} = (-C/2D)^{1/2}$ (in unit of $\ell^{-1}$) may
dominate over the two other phases.
(It should be noted that $C<0$ is only a necessary condition to have
the rippled phase as will be discussed below.)
To obtain the mean-field phase diagram, we employ the single-mode
approximation in which the rippled phase is described  by the most 
dominant wavevector $q^{\ast}$, only.
This approximation is valid near the phase transition point where the
segregation tendency between the two lipids is weak enough.
The order parameter $m$ can be written as \cite{MFLM}
\begin{equation}
m=m_0 + m_1 \cos(q^{\ast} x),
\label{order_parameter}
\end{equation}
where $m_0=\langle m \rangle$ is the spatial average of $m$, and $m_1$ 
is the amplitude of the single $q^{\ast}$-mode of the rippled phase 
taken to lie arbitrarily in the $x$-direction.

\begin{figure}
\begin{center}
\resizebox{0.45\textwidth}{!}{%
\includegraphics{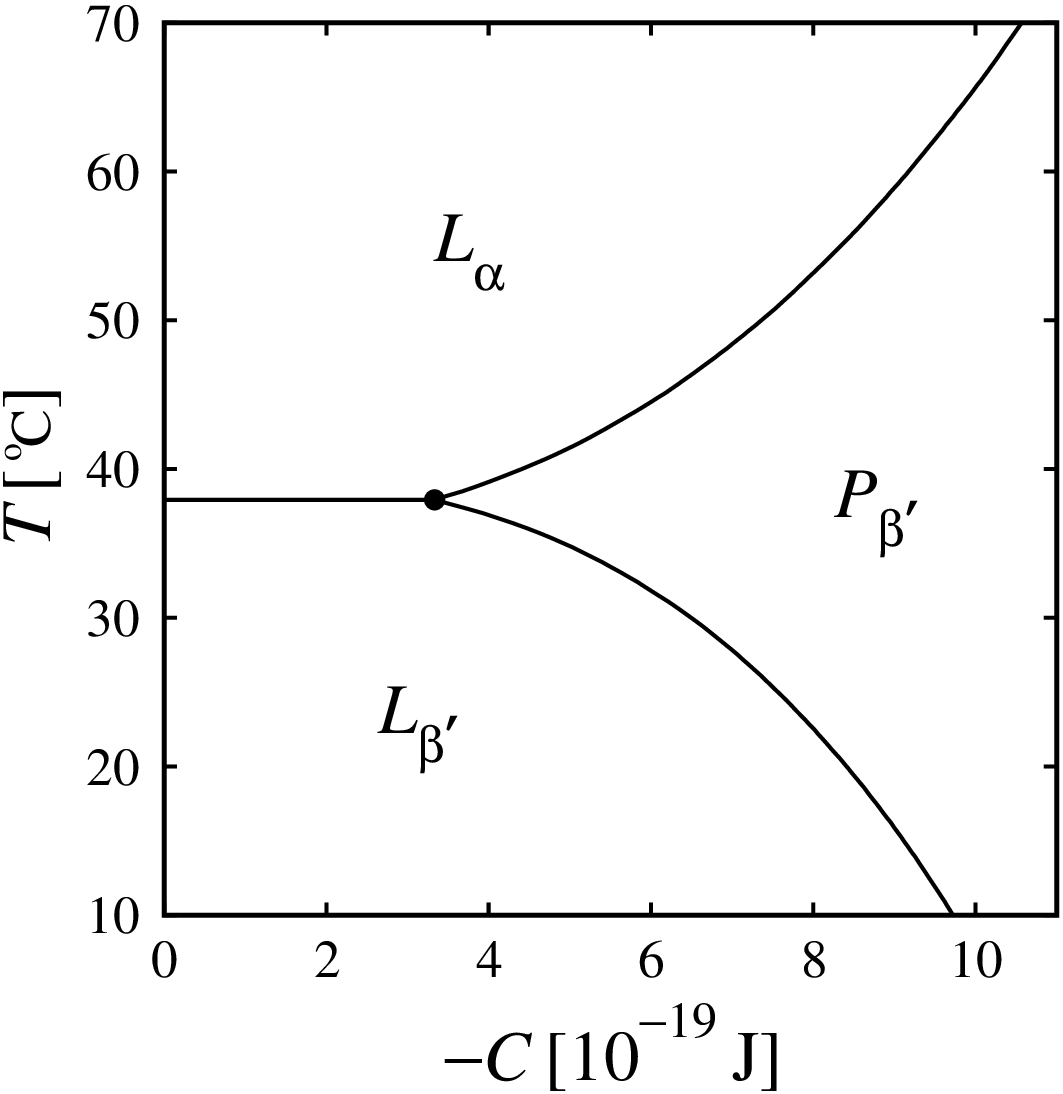}
}
\end{center}
\caption{Mean-field phase diagram of a single-component
lipid membrane as a function of (negative) elastic constant
$C$ and temperature $T$ (in degrees Celsius).
The liquid ($L_{\alpha}$), solid ($L_{\beta'}$) and rippled
($P_{\beta'}$) phases are separated from one another by first-order
phase transition lines which meet at a triple point
(marked by a full circle):
$C_{\rm tr}=-3.33 \times 10^{-19}$\,J,
$T_{\rm tr}=38$\,\textdegree C.
The parameters used to calculate this phase diagram are given
in the text.
}
\label{fig1}
\end{figure}

By substituting Eq.~(\ref{order_parameter}) into
Eq.~(\ref{free_energy}) and taking the spatial average over
one period, we obtain
\begin{eqnarray}
\langle f_{\rm st} \rangle & = & - \frac{C^{2}}{16D}m_1^2 +
\frac{1}{2}a_2  \left(m_0^2+\frac{1}{2}m_1^2\right)
\nonumber  \\
& & +
\frac{1}{3}a_3 \left(m_0^3 + \frac{3}{2}m_0 m_1^2\right)
\nonumber \\
& & + \frac{1}{4}a_4
\left(m_0^4 + 3 m_0^2 m_1^2 + \frac{3}{8} m_1^4\right).
\label{mean_field}
\end{eqnarray}
The two homogeneous liquid and solid phases, $L_{\alpha}$ and
$L_{\beta'}$, respectively, are characterized by $m_1 = 0$; while
$m_0=0$ in the disordered liquid $L_{\alpha}$ phase, and is
non-zero in the solid  $L_{\beta'}$ phase. 
From the three phases only the $P_{\beta'}$ phase is characterized 
by a modulating amplitude, $m_1 \neq 0$.

The mean-field phase diagram is obtained by minimizing
Eq.~(\ref{mean_field}) with respect to both $m_0$ and $m_1$, and
comparing the relative stability of the three relevant phases. 
In Fig.~\ref{fig1}, we present an example of such a phase diagram,
as function of the (negative) effective elastic constant $C$ and
temperature $T$ plotted in degrees Celsius.
The parameter $C$ can be a function of relative humidity (fraction 
of water content in the lamellar phase) as discussed below. 
The phase diagram is calculated for a choice of system
parameters which reproduce the main transition temperature of 
DMPC at $T_{\rm m}=38$\,\textdegree C:
$a'_2 = 2.4 \times 10^{-21}$\,J\,K$^{-1}$,
$a_3 = -1.1 \times 10^{-18}$\,J,
$a_4= 2.2 \times 10^{-18}$\,J,
$T^{\ast}=-13$\,\textdegree C ($260$\,K), and 
$D= 2.0 \times 10^{-18}$\,J.
These values (except $D>0$ whose value is less important) are taken 
from Ref.~\cite{GL2}.
By inserting the above Landau coefficients into Eq.~(\ref{main}), 
it can be readily checked that the main transition temperature of 
DMPC $T_{\rm m}=38$\,\textdegree C  (Fig.~\ref{fig1}) is obtained.

In Fig.~\ref{fig1}, three first-order phase transition lines separate 
the three phases and meet at a triple point,
$C_{\rm tr} = -3.33 \times 10^{-19}$\,J.
Due to the first-order nature of the $L_{\beta'} \rightarrow
L_{\alpha}$ phase transition, it can be argued, on general grounds, 
that the triple point $C_{\rm tr}$ has to be located at non-zero values
of $C$.
For small magnitude of $C$, $\vert C \vert < \vert C_{\rm tr} \vert$,
the horizontal first-order phase transition line occurs at
$T=T_{\rm m}=38$\,\textdegree C and is $C$ independent.
An increase in temperature (as long as 
$ \vert C\vert <\vert C_{\rm tr}\vert$) will melt the solid phase
$L_{\beta'}$ directly into the liquid phase $L_{\alpha}$ at $T_{\rm m}$.
But for negative and large enough magnitude of $C$,
$\vert C \vert > \vert C_{\rm tr} \vert$, any increase in temperature
will cause a sequence of phase transitions: first the solid phase 
$L_{\beta'}$ melts into the $P_{\beta'}$ phase, and only then, upon 
further increase of the temperature, the $P_{\beta'}$ phase will make 
a phase transition into the liquid phase $L_{\alpha}$.
It is also apparent from the figure that the region of the $P_{\beta'}$
phase expands on the expense of the uniform liquid and solid phases as
$\vert C \vert$ increases.
This means that the phase transitions $L_{\beta'} \rightarrow
P_{\beta'}$ and $P_{\beta'} \rightarrow L_{\alpha}$ (at constant $C$)
occur at larger temperature deviations from $T_{\rm m}$ as
$\vert C \vert$ increases.

The phase diagram of Fig.~\ref{fig1} essentially reproduces
all the experimental facts observed for single-component lipid
membranes.
It qualitatively agrees with the observed phase diagram of a lamellar
phase of DMPC, when the negative elastic constant $C$ is taken to be 
proportional to the relative humidity \cite{SSSPC,SSSC}.
Although the exact dependence of the elastic constant $C$
on humidity is not known, the former can possibly be reduced
by increasing hydration.
For another lipid, DPPC, the $P_{\beta'}$ phase is found experimentally
to occur at temperatures around $37$\,\textdegree C 
(in Ref.~\cite{BWGG} the precise temperatures for the $L_{\beta'} 
\rightarrow P_{\beta'}$ and $P_{\beta'} \rightarrow L_{\alpha}$ phase 
transitions have not been fully reported), while the rippled phase is 
not observed for a stack DPPE lipid bilayers.
Such a difference in the lipid phase behavior can be qualitatively
attributed to  different values of the elastic constant $C$.

\section{Two-component lipid mixtures}
\label{sec:binary_model}

\begin{figure}
\begin{center}
\resizebox{0.45\textwidth}{!}{%
\includegraphics{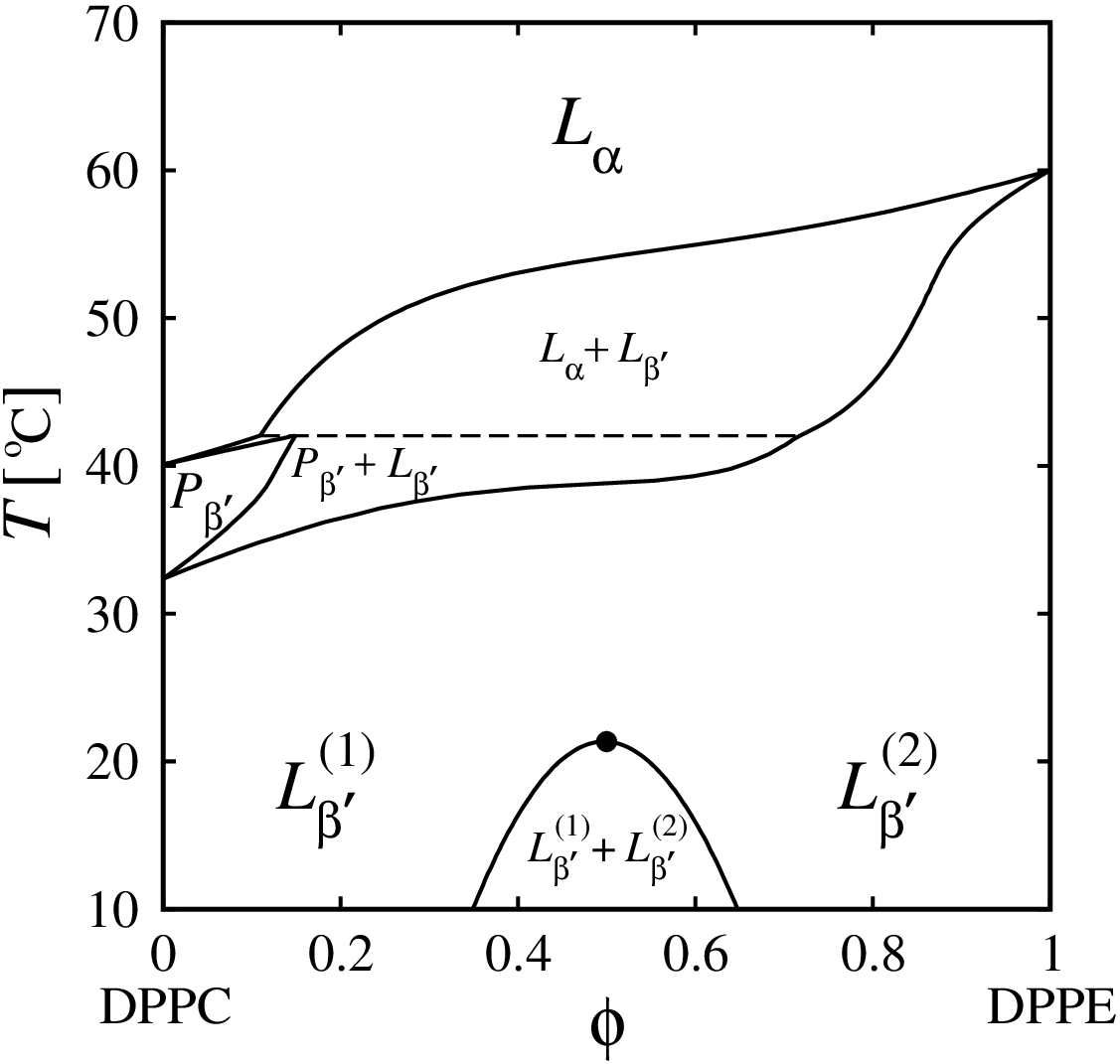}
}
\end{center}
\caption{Calculated mean-field phase diagram of a binary lipid mixture
as a function of their relative composition $\phi$ and temperature $T$.
The parameter values are chosen to fit DPPC/DPPE mixtures
($\phi=0$ represents pure DPPC and is denoted by a subscript ``B''):
$C_{\rm A}=-2.0 \times 10^{-19}$\,J,
$C_{\rm B}=-5.2 \times 10^{-19}$\,J,
$T^{\ast}_{\rm A}=9$\,\textdegree C,
$T^{\ast}_{\rm B}=-15$\,\textdegree C, 
and $J=1.45\times 10^{-20}$\,J.
All other parameter values and definitions of the different phases
are the same as in Fig.~\ref{fig1}.
The critical point is indicated by a full circle and occurs at
$T_{\rm c}=21.3$\,\textdegree C.
The horizontal dashed line indicates the three-phase coexistence
at the triple point, $T_{\rm tr}=42.0$\,\textdegree C.
}
\label{fig2}
\end{figure}

\begin{figure}
\begin{center}
\resizebox{0.45\textwidth}{!}{%
\includegraphics{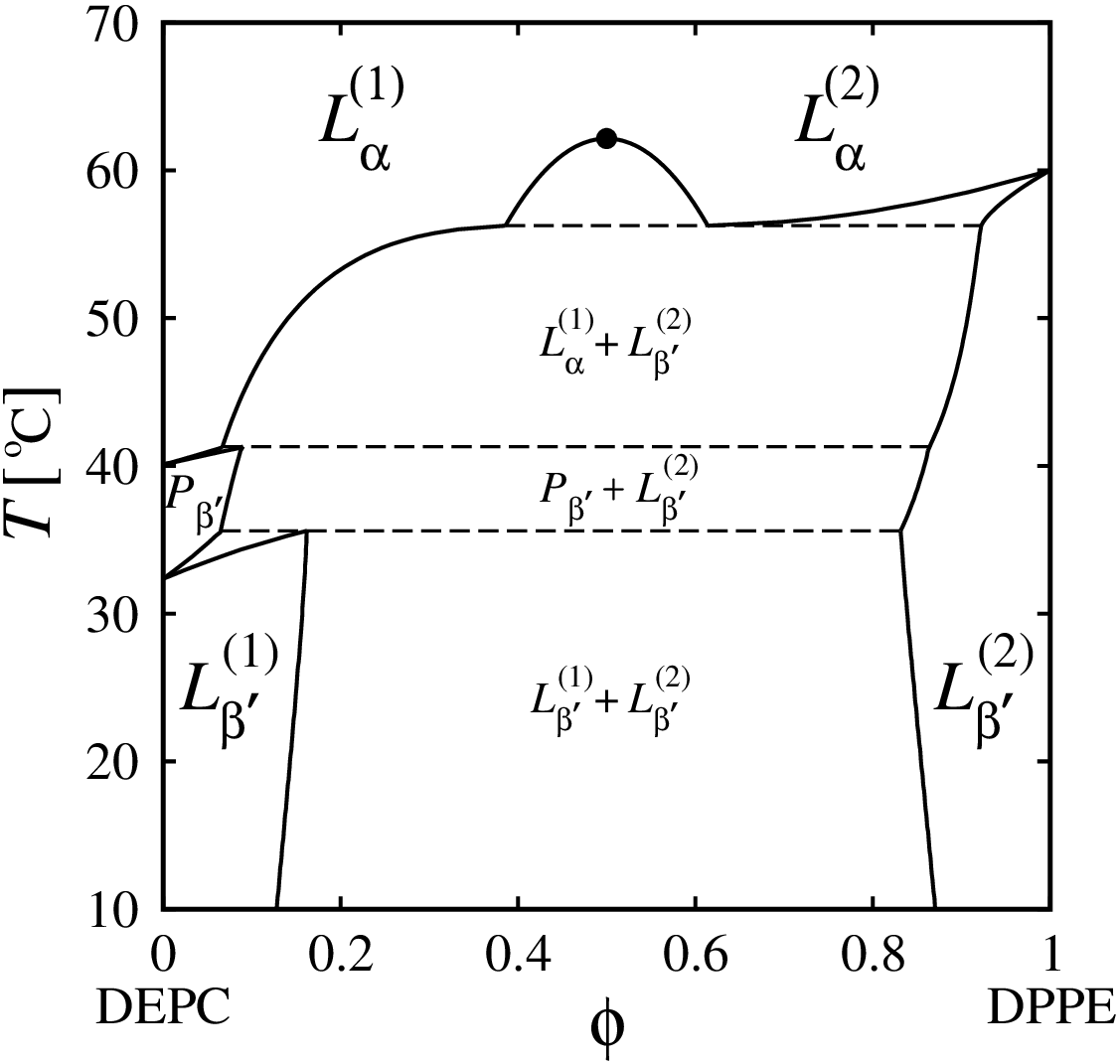}
}
\end{center}
\caption{Calculated mean-field phase diagram of a binary lipid mixture
as a function of their relative composition $\phi$ and temperature $T$.
The parameters are the same as in Fig.~\ref{fig2} except
$J=1.85 \times 10^{-20}$\,J.
The critical point is located at $62.1$\,\textdegree C, and
three triple points occur at 35.6, 41.3, and 56.3\,\textdegree C, 
respectively.
This choice of model parameters is in agreement with experimental
phase diagram of DEPC/DPPE mixture where $\phi=0$ represents pure DEPC.
}
\label{fig3}
\end{figure}

In this section, we extend the above single-component model to
membranes consisting of a binary lipid mixture.
We employ  a similar approach as was used in our previous study
\cite{KSOA,KSO}, and consider the coupling of the melting phase
transition with a lateral phase separation in the mixed
membrane.

A binary mixed membrane is modeled as an incompressible
A/B mixture of $\phi$ mole
fraction of lipid A and $(1-\phi)$ fraction of lipid B.
For simplicity, we assume the same area per molecule for both species
and ignore any lipid exchange with the surrounding solvent.
In general, the two lipids will have different main transition
temperatures originating from different molecular parameters such as
chain length, degree of saturation and hydrophilic head group.

The total free energy per lipid, $f_{\rm tot}=f_{\rm st}+f_{\rm mix}$,
comprises: (i) the chain stretching free energy $f_{\rm st}$
given by Eq.~(\ref{free_energy}), and (ii) the free energy
of mixing, $f_{\rm mix}$.
The latter energy per lipid molecule is the sum of the entropy of
mixing and enthalpy.
It can be written within the Bragg-Williams (mean-field) approximation 
as
\begin{equation}
f_{\rm mix} = k_{\rm B} T [\phi \log \phi + (1-\phi) \log (1-\phi)] + 
\frac{1}{2}J  \phi (1-\phi),
\label{phase_separation}
\end{equation}
where $k_{\rm B}$ is the Boltzmann constant, and $J>0$ is an attractive
interaction parameter between the lipids that enhances lipid-lipid
demixing.

For a binary mixture the free energy $f_{\rm st}$ is assumed to have
the same functional dependence on the effective elastic constant $C$
and the reference temperature $T^{\ast}$.
Although these two parameters depend on the lipid composition $\phi$,
the precise dependence cannot be calculated from such a phenomenological 
approach. 
Alternatively, we proceed by further assuming the simplest linear 
interpolation between the two pure lipid limiting values:
\begin{equation}
C(\phi)=\phi C_{\rm A} +(1-\phi)C_{\rm B},
\label{elastic_constant}
\end{equation}
\begin{equation}
T^{\ast}(\phi) = \phi T^{\ast}_{\rm A} +
(1-\phi)T^{\ast}_{\rm B},
\label{temperature}
\end{equation}
where $C_{\rm A}$ and $C_{\rm B}$ are the elastic constants,
and $T^{\ast}_{\rm A}$ and $T^{\ast}_{\rm B}$ are the reference
temperatures of lipid A and B, respectively.
The other elastic constant $D$, as well as $a_3$ and $a_4$ in
Eq.~(\ref{free_energy}) are assumed to be the same for the two
lipids because the important properties of the lipids are mostly
reflected in the choice of their $C$ and $T^{\ast}$ parameter values.
As for all other model parameters, we use hereafter the
same values listed in the previous section.

The total free energy $f_{\rm tot}=f_{\rm st}+f_{\rm mix}$ as function
of $m$ and $\phi$ is now readily available by substituting
Eqs.~(\ref{elastic_constant}) and (\ref{temperature}) into
Eq.~(\ref{free_energy}).
There is an explicit coupling between the two order parameters,
$m$ and $\phi$ originating from  the free energy
$f_{\rm st}=f_{\rm st}(m,\phi)$.
Note that the first term in Eq.~(\ref{free_energy}) and 
Eq.~(\ref{temperature}) lead to a coupling term which scales as 
$\phi m^2$. 
Another possible coupling term is $\phi^2 m$ that simply renormalizes
the interaction parameter $J$ of $f_{\rm mix}$ and is less interesting.
The two-phase coexistence region in the ($T$, $\phi$) plane is
calculated by using the common tangent construction to account for 
the constraint value of the relative concentration $\phi$, after 
minimizing $f_{\rm tot}$ with respect to $m_0$ and $m_1$.

By properly choosing the model parameters, we attempt to reproduce
the binary phase diagram of several lipid mixtures, the first being
that of DPPC/DPPE~\cite{BWGG} shown in Fig.~\ref{fig2}.
Since a pure DPPE membrane does not show experimentally the rippled phase,
its elastic constant is chosen as $C_{\rm A}=-2.0 \times 10^{-19}$\,J.
Namely, below the triple point $C_{\rm tr}$ (Fig.~\ref{fig1}).
The DPPE reference temperature $T_{\rm A}^*$ is calculated from 
Eq.~(\ref{main}) with the same values for the Landau 
coefficients as before, by fitting its main transition temperature
to the observed one of  $T_{\rm m}=60$\,\textdegree C.
On the other hand, the elastic constant of DPPC is chosen as
$C_{\rm B}=-5.2 \times 10^{-19}$\,J (above $C_{\rm tr}$) so that
the $P_{\beta'}$ phase appears between the $L_{\beta'}$ and 
$L_{\alpha}$ phases.
The reference temperature of DPPC is set as
$T^{\ast}_{\rm B}=-15$\,\textdegree C,
corresponding, from Eq.~(\ref{main}), to a $L_{\beta'} \rightarrow 
L_{\alpha}$ transition temperature $T_{\rm m}=36$\,\textdegree C.
This temperature is not available from experiments but lies in-between 
the experimentally observed $L_{\beta'} \rightarrow P_{\beta'}$ phase 
transition temperature at $32$\,\textdegree C and the $P_{\beta'} 
\rightarrow L_{\alpha}$ one at $40$\,\textdegree C.

All the above chosen parameter values are used to calculate
the binary phase diagram fitting the DPPC/DPPE mixture.
The relative composition $\phi$ is chosen such that $\phi=0$
corresponds to pure DPPC, while $\phi=1$ to pure DPPE.
In addition, the interaction parameter is chosen as
$J=1.45 \times 10^{-20}$\,J.
When the temperature is relatively low, there is a coexistence
region between two $L_{\beta'}$ phases (denoted by
$L_{\beta'}^{(1)}+L_{\beta'}^{(2)}$).
This low temperature coexistence terminates at a critical temperature,
above which there is only one $L_{\beta'}$ phase for the entire range
of composition $\phi$. 
It should be noted, however, that the appearance of the critical point 
between the two ``solid'' phases is an artifact of the model.
As the temperature is further increased, the $P_{\beta'}$ phase
appears close to the pure DPPC axis.
It is bound by a coexistence region with the $L_{\beta'}$ phases
as well as with another (much smaller in its extent)
coexistence region with the $L_{\alpha}$ phase. 
This latter region is very small on the
scale of the figure and collapses almost into a line.
These two coexistence regions terminate at a triple point temperature,
$T=T_{\rm tr}$, at which all the three phases coexist.
For even higher temperature, $T>T_{\rm tr}$, the rippled phase
disappears and the only coexistence region is between the
$L_{\alpha}$ and $L_{\beta'}$ phases.

The calculated phase diagram is in quantitative agreement
with the experimental one for  DPPC/DPPE
mixtures (e.g., Fig.~7 of Ref.~\cite{BWGG}).
Even the coexistence between the two solid phases is alluded
in Ref.~\cite{BWGG}.
A similar type of phase diagram was obtained for mixtures
of DMPC/DPPS and DPPC/DPPS \cite{LMcC}.
In these  lipid mixtures, both DMPC and DPPC exhibit
the rippled phase, while pure DPPS undergoes a direct transition
from the $L_{\beta'}$ to $L_{\alpha}$ phases similar to DPPE.

\begin{figure}
\begin{center}
\resizebox{0.45\textwidth}{!}{%
\includegraphics{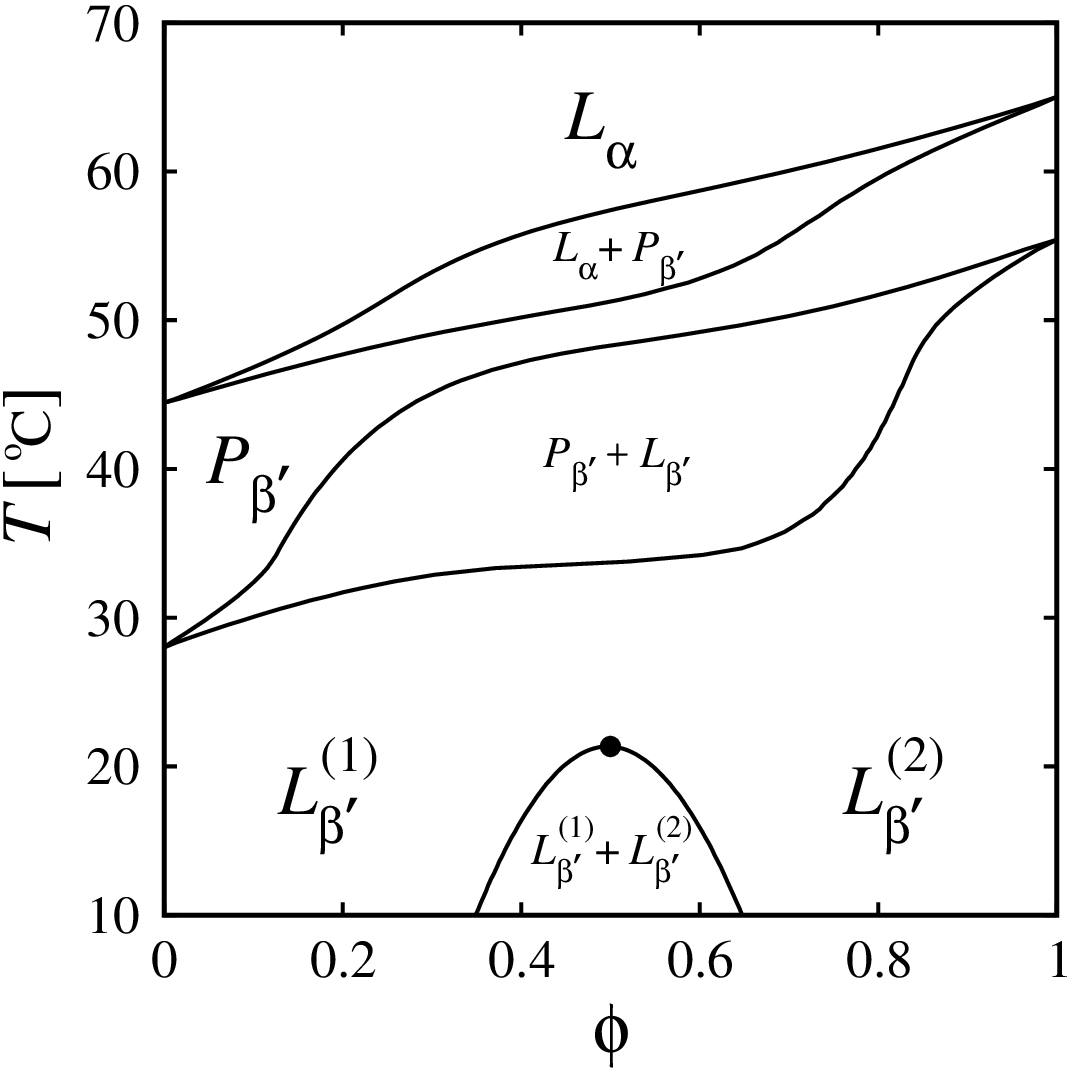}
}
\end{center}
\caption{Mean-field phase diagram of a binary mixture of two lipids
as a function of the relative composition $\phi$ and temperature $T$.
All parameters are the same as in Fig.~\ref{fig2} except
$C_{\rm A}=-5.5 \times 10^{-19}$\,J and
$C_{\rm B}=-6.5 \times 10^{-19}$\,J.
The lipid interaction parameter is set to be $J=1.45 \times 10^{-20}$\,J.
The critical point is located at $21.3$\,\textdegree C as in
Fig.~\ref{fig2}.
}
\label{fig4}
\end{figure}

\begin{figure}
\begin{center}
\resizebox{0.45\textwidth}{!}{%
\includegraphics{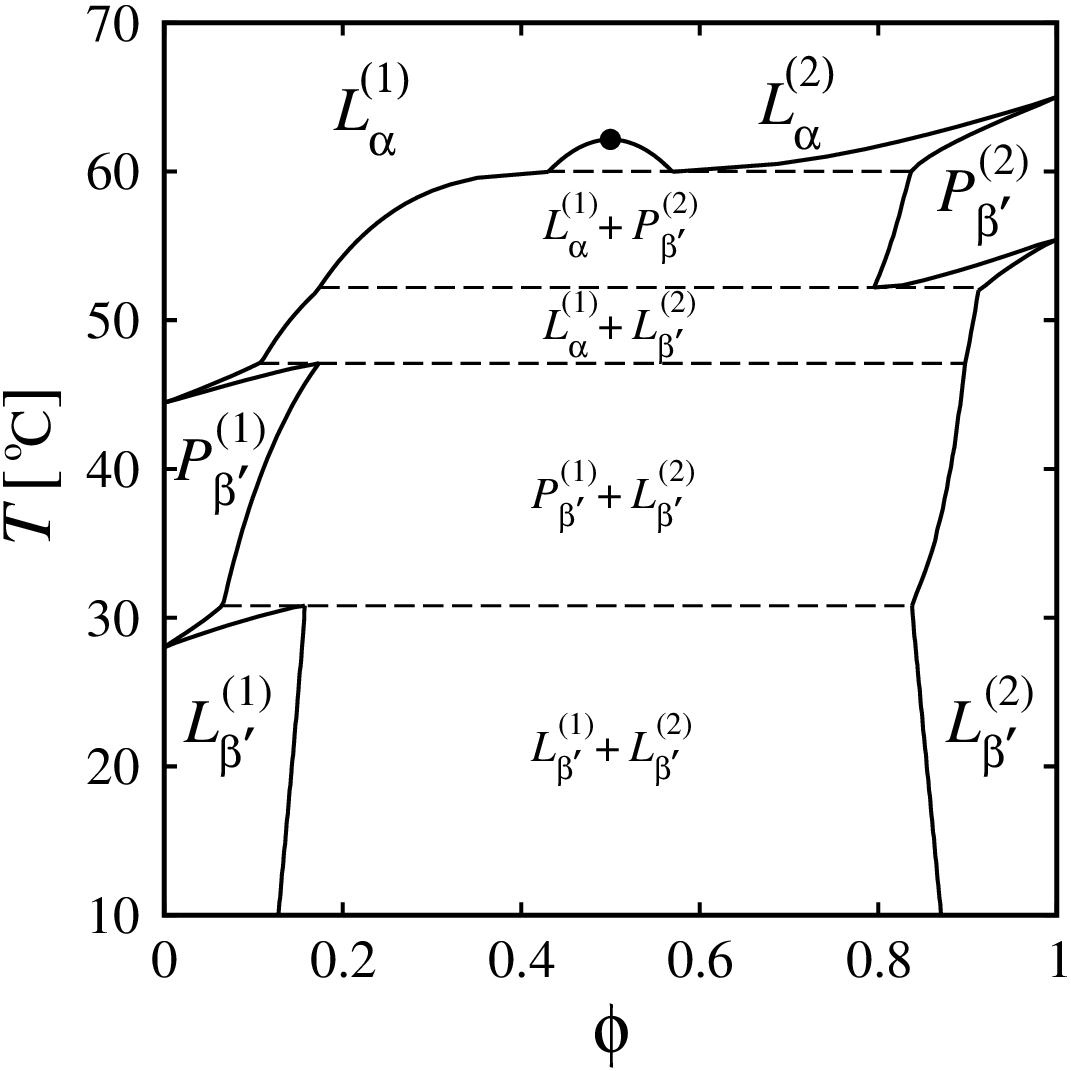}
}
\end{center}
\caption{Mean-field phase diagram of a binary mixture of two lipids
as a function of the relative composition $\phi$ and temperature $T$.
All parameters are the same as in Fig.~\ref{fig4} except
$J=1.85 \times 10^{-20}$\,J.
The critical point is located at $62.1$\,\textdegree C as in
Fig.~\ref{fig3}, and four triple points occur at 30.8, 47.1, 52.2,
and 60.0\,\textdegree C,
respectively.
}
\label{fig5}
\end{figure}

An even stronger lipid-lipid segregation can be modeled by increasing 
the value of the lipid interaction parameter to 
$J=1.85 \times 10^{-20}$\,J, while keeping all other parameter values
unchanged. 
In the resulting phase diagram shown in Fig.~\ref{fig3}, the lower 
coexistence region between the two $L_{\beta'}$ phases penetrates 
the other coexistence regions.  
Such a phase diagram resembles that of DEPC/DPPE lipid mixtures 
\cite{WM}.
This combination of lipids have a strong segregation tendency
because both head and tail moieties are different. 
In the calculated phase diagram, there are three triple points
indicated by the horizontal dashed lines.
The rippled phase can now coexist with either of the $L_{\beta'}$
(regions $P_{\beta'}+L_{\beta'}^{(1)}$ or $P_{\beta'}+L_{\beta'}^{(2)}$
in the figure) as well as with the $L_{\alpha}$ phase
($P_{\beta'}+L_{\alpha}^{(1)}$).
There is also a two-phase coexistence region
between the two liquid phases ($L_{\alpha}^{(1)}+L_{\alpha}^{(2)}$).
The three-phase coexistence between the two $L_{\alpha}$ phases
and the $L_{\beta'}$, and the existence of the critical point
have been indeed observed for DEPC/DPPE lipid mixtures \cite{WM}.

In Figs.~\ref{fig4} and \ref{fig5} we calculate the phase diagrams
in situations where each of the two pure lipids exhibit a rippled 
phase between their corresponding $L_{\alpha}$ and $L_{\beta'}$ phases.
Hence, we choose the magnitudes of the elastic constant to be larger
than $\vert C_{\rm tr}\vert$
of Fig.~\ref{fig1}:
$C_{\rm A}=-5.5 \times 10^{-19}$\,J and
$C_{\rm B}=-6.5 \times 10^{-19}$\,J.
The choices of the reference temperatures $T^{\ast}_{\rm A}$ and
$T^{\ast}_{\rm B}$, as well as all other parameters, is the same as
in Figs.~\ref{fig2} and \ref{fig3}.
The interaction parameter is taken as $J=1.45 \times 10^{-20}$\,J
in Fig.~\ref{fig4}, and have a somewhat higher value, 
$J=1.85 \times 10^{-20}$\,J, in Fig.~\ref{fig5}.
In the  phase diagram of Fig.~\ref{fig4}, one region of the rippled
phase extends throughout the entire range of the lipid composition.
Above and below this region, the rippled phase coexists with the
$L_{\beta'}$ and $L_{\alpha}$ phases.
At low temperatures, another coexistence region can be seen between
the two $L_{\beta'}$ phases.
For a larger interaction parameter $J$ (Fig.~\ref{fig5}), these
coexistence regions merge to form a large two-phase region, and there
are four triple points.
Here we see a new coexistence between $L_{\alpha}$ and $L_{\beta'}^{(2)}$
phases which does not exist in Fig.~\ref{fig4}.
Although such more complex phase diagrams have not  been yet reported
in experiments, we expect that they can be found in the future by 
properly choosing the lipid mixtures.
One possible choice of  lipids that may give rise to such a
phase behavior is the DPPC/DMPC mixture. Both lipids exhibit the
rippled phase between their $L_{\beta'}$ and $L_{\alpha}$ phases.

\section{Summary and discussion}
\label{sec:discussion}

In this paper we proposed a model describing the liquid-solid
coexistence region in membranes composed of a binary lipid mixture. 
We addressed in particular the possible existence of a rippled phase 
between the liquid and solid ones, and its effect on the global phase 
diagrams.
In order to incorporate the possibility of a rippled phase, we use a
model previously proposed for  single-component membranes by Goldstein 
and Leibler (GL) \cite{GL1,GL2}.
Their main purpose was to present a model describing the lyotropic
lamellar phases of lipid bilayers as stacks of \textit{interacting}
membranes.
Here we have shown explicitly that the GL lipid stretching energy
of an \textit{isolated} membrane exhibits the sequential phase 
transitions $L_{\beta'} \rightarrow P_{\beta'} \rightarrow
L_{\alpha}$ by increasing the temperature.
We further extend the single-component model  and apply it to binary
lipid mixtures  assuming that  the elastic parameter $C$ as well as the
reference temperature $T^{\ast}$ depend linearly on the relative 
composition of the two lipids.
The calculated phase diagrams are in quantitative agreement with the
experimental ones for specific lipid mixtures.
We have also predicted other types of phase diagrams,
which have yet
to be checked experimentally for other lipid mixtures.

Several points merit further discussion.
In Sec.~\ref{sec:single}, we choose the relative membrane thickness
as the scalar order parameter to describe the membrane phase transitions
\cite{GL1,GL2}.
In another model proposed by Chen, Lubensky and MacKintosh (CLM),
a two-dimensional molecular tilt was used as a vector order
parameter \cite{LM,CLM}.
In their model, a coupling mechanism between the membrane curvature
and the gradient in molecular tilt was considered.
As a result, various types of rippled phases with different
inplane symmetries have been predicted.

In spite of the success in predicting, e.g., the square
lattice phase that was found in the experiments \cite{YF}, the
CLM model suffers from some deficiencies which makes it inappropriate
to be used in our study.
Within the CLM model, the $L_{\beta'} \rightarrow P_{\beta'}$
phase transition cannot be induced only by changing the temperature,
while keeping all other system parameters fixed. In other words,
the phase boundaries in the CLM model is parallel to the
temperature axis (see, for example, Fig.~9 in Ref.~\cite{CLM}).
This is essentially a result of the
vector nature of the CLM order parameter \cite{CL}.
Unfortunately, such a phase behavior is not in accord with the
experimental phase diagrams \cite{SSSPC,SSSC}, where the
$L_{\beta'}$\,--\,$P_{\beta'}$ phase boundary depends on
temperature.
This is the main reason why we have chosen to use a scalar order 
parameter giving us the phase diagram as in Fig.~\ref{fig1}.

The second problem is that the CLM model predicts a second-order
phase transition between the $L_{\alpha}$ and  $L_{\beta'}$ phases.
However, the main transition is known to be first-order
according to various experimental investigations \cite{GL1,GL2},
as was previously mentioned in Ref.~\cite{YF}.
A possible way out would be to include the 6th order term of the
tilting vector in the CLM Landau expansion, this lies beyond the 
scope of the present work.

The two problems  mentioned above do not exist in the GL model.
However, since the choice of a single scalar as the
order parameter is a major simplification, many degrees of freedom
of the lipid molecules are not properly taken into account.
For example, one cannot distinguish between the similar $L_{\beta'}$ and
$L_{\beta}$ phases because the scalar order parameter $m$ cannot
model the molecular tilt characteristic of the solid-like $L_{\beta}$ 
phase.
With this choice of order parameter, one can only distinguish between 
a dilute and a condensed phase.
Because the membrane thickness does not account for the different 
symmetries of the $L_{\alpha}$, $L_{\beta'}$, and $P_{\beta'}$ phases,
$m$ cannot represent any crystalline order, and the solid phase is 
not properly described here.  
Hence the appearance of the critical points between the two 
$L_{\beta'}$ phases in Figs.~\ref{fig2} and \ref{fig4} is an 
artifact of our model. 
Although this problem is remedied  in the CLM model,
the GL model captures the essential
features of the structural transitions in lipid membranes in the
presence of the rippled phase.
Using the latter model is sufficient to reproduce even quantitatively
some of the global features of  phase diagrams
of binary lipid mixtures, in agreement with experiments.

In Eq.~(\ref{free_energy}) we  assumed for simplicity that the rippled
phase is modulated  spatially only in the $x$-direction.
In general, one can  consider the full $(x,y)$ inplane modulations
as treated in the CLM model \cite{LM,CLM}.
This can be accomplished by including in Eq.~(\ref{free_energy}) 
additional gradient terms in the $y$-direction.
The resulting phase diagram will be more complex because there
can be more than one rippled phase.
Although a two-dimensional square lattice phase has been identified
for a single-component DTPC lipid \cite{YF}, it has not yet been
found in binary lipid mixtures.
We also note that the one-dimensional modulation of the order
parameter expressed by Eq.~(\ref{order_parameter}) is similar to
the $P_{\beta'}^{(2)}$ phase in CLM,  with the main difference that
we include a non-zero average term $m_0$.
The physical interpretation of the actual order parameter in the
two models differs as is explained above.

Finally, we discuss why the elastic coefficient $C$ can take 
negative values in Eq.~(\ref{free_energy}) as is required to get 
the $P_{\beta'}$ phase at equilibrium. 
In Sec.~\ref{sec:single}, we have mentioned that the coupling between
the conformation of the chains and the curvature of the membrane may
be a possible reason \cite{GL1}.
In general, if there exists a coupling between an additional elastic
degree of freedom and the gradient of the order parameter, the
elastic constant $C$ will be reduced.
This situation can be expressed by including the following two
additional terms in the free energy:
\begin{equation}
f_{\rm el} = \frac{1}{2}B \rho^2 - \gamma \rho
\left( \frac{{\rm d} m}{{\rm d} x} \right).
\label{eq:coupling}
\end{equation}
Here $\rho$ represents some elastic degree of freedom, $B$ is another
positive elastic constant, and $\gamma$ is a coupling constant.
For example, $\rho$ can be taken as the curvature of the lipid/water
interface, and $B$ is the bending rigidity of the membrane.
In the second term, $\rho$ is linearly coupled to the gradient of $m$
which induces a spontaneous value of $\rho$.
Minimizing $f_{\rm el}$ of Eq.~(\ref{eq:coupling}) with respect to
$\rho$, we get 
\begin{equation}
\rho = \frac{\gamma}{B} \left( \frac{{\rm d} m}{{\rm d} x} \right).
\label{eq:minimum}
\end{equation}
This means that the equilibrium value of $\rho$ is proportional to the
gradient of $m$.
By substituting this equation back into Eq.~(\ref{eq:coupling}),
the effective elastic constant becomes
\begin{equation}
C \rightarrow C - \gamma^2/B,
\label{eq:constant}
\end{equation}
in Eq.~(\ref{free_energy}). 
Hence $C$ is reduced by $\gamma^2/B$ and can reach negative values
when the coupling constant $\gamma$ is large enough (or when the 
elastic constant $B$ is small enough).
Although it goes beyond the scope of the present paper to suggest 
specific molecular mechanism to account for this added elastic degree 
of freedom, such a mechanism would, in principle, explain the negative 
value of the effective elastic constant, $C$.


\begin{acknowledgement}
This work is supported by KAKENHI (Grant-in-Aid for Scientific
Research) on Priority Areas ``Soft Matter Physics'' and
Grant no.\ 18540410 from the Ministry of Education, Culture, Sports,
Science and Technology of Japan.
One of us (DA) acknowledges support from the Israel Science Foundation
(ISF) under grant no.\ 160/05 and the US-Israel Binational Foundation
(BSF) under grant no.\ 287/02.
\end{acknowledgement}


\end{document}